\documentclass[reprint,twocolumn,secnumarabic,amssymb, nobibnotes, aps, pra,superscriptaddress]{revtex4-1}

\usepackage{extarrows}
\usepackage{amsmath}    
\usepackage{graphicx}    
\usepackage{verbatim}    
\usepackage{color}           
\usepackage{subfigure}   
\usepackage{braket}
\usepackage{hyperref}    
\usepackage{verbatim}  
\usepackage{here}
\usepackage{mathrsfs}
\usepackage[english]{babel}
\usepackage[utf8x]{inputenc}
\usepackage[T1]{fontenc}
\usepackage{comment}

\graphicspath{{Figures/}}
\usepackage{physics}
\usepackage{siunitx}
\usepackage{bm}
\usepackage[colorinlistoftodos]{todonotes}

\newcommand{\bR}{\mathbf{R}}
\newcommand{\bk}{\mathbf{k}}

\newcommand{\bphi}{\bm{\phi}}

\definecolor{blue(pigment)}{rgb}{0.2, 0.2, 0.6}
\definecolor{blue-violet}{rgb}{0.54, 0.17, 0.89}

\begin{document}
\title{Dynamical response and competing orders in two-band Hubbard model}

\author{A. Niyazi}
\affiliation{Institute of Solid State Physics, TU Wien, 1040 Vienna, Austria}
\author{D. Geffroy}
\affiliation{Department of Condensed Matter Physics, Faculty of
  Science, Masaryk University, Kotl\'a\v{r}sk\'a 2, 611 37 Brno,
  Czechia}
\affiliation{Institute of Solid State Physics, TU Wien, 1040 Vienna, Austria}
\author{J. Kune\v{s}}
\affiliation{Institute of Solid State Physics, TU Wien, 1040 Vienna, Austria}
\affiliation{Institute of Physics, Czech Academy of Sciences, Na Slovance 2, 182 21 Praha 8, Czechia}
\date{\today}

\begin{abstract}
We present a dynamical mean-field study of two-particle dynamical response functions in two-band
Hubbard model across several phase transitions. We observe that the transition between the excitonic
condensate and spin-state ordered state is continuous with a narrow strip of supersolid phase separating the two.
Approaching transition from the excitonic condensate is announced by softening of the excitonic mode at 
the $M$ point of the Brillouin zone. Inside the spin-state ordered phase there is a magnetically ordered
state with $2\times2$ periodicity, which has no precursor in the normal phase. 
\end{abstract}

\maketitle

\section{Introduction}
Spontaneous symmetry breaking, which accompanies the continuous phase
transitions, changes qualitatively the dynamical response of solids. 
If the broken symmetry is continuous, low-energy Goldstone mode(s)
associated with the long-wavelength dynamics of the order parameter,
appears in systems with short-range interactions.
Excitonic condensates (ECs)~\cite{Mott1961,Keldysh1965,Halperin1968b}
represent an exotic type of broken-symmetry phase. While the
experimental realizations of EC has been limited to artificial
structures such as quantum wells in strong magnetic
field~\cite{Eisenstein2004} or cavity systems~\cite{Balili2007},
recent experiments on 1T-TiSe$_2$~\cite{Cercellier2007,Kogar2017} ,
Ca$_2$RuO$_4$~\cite{Jain2017} or
Pr$_{0.5}$Ca$_{0.5}$CoO$_3$~\cite{tsubouchi2002,Moyoshi2018} revived
the interest in the subject also in bulk solids. Condensation of
spinful excitons, which gives rise to a new type of magnetic behavior
is particularly interesting. The simplest model to capture the
excitonic magnetism is the two-orbital Hubbard model at half
filling~\cite{Kunes2014b,Hoshino2016,Kaneko2014} and its
strong-coupling limit~\cite{Khaliullin2013,Nasu2016,Tatsuno2016}. The
parameter range of interest hosts a number of ordered
phases~\cite{Kunes2015,Nasu2016} in addition to the first-order
metal-insulator transition~\cite{Werner2007}. Besides the general
interest in understanding its behavior, the model provides a fertile
playground for testing theoretical methods.

Computation of two-particle (2P) response for
realistic materials is a challenging task. Dynamical
mean-field theory (DMFT)~\cite{Georges1996,Kotliar2006} has been
successful in bringing together the material realism of multi-orbital
models with the many-body realism, including real temperatures, phase
transitions, quasi-particle life times, atomic-multiplet effects. Despite the boom of the past two decades, application of DMFT has
been largely limited to one-particle (1P) quantities, such as
generalized band structures and occupation numbers. Solved in principle,
the calculation of 2P response functions is numerically very demanding 
as it involves the solution of the Bethe-Salpeter equation for large multi-index
objects. There are compelling reasons to study the 2P response within
DMFT. Most experimental probes and applications employ the 2P response 
of materials. Current density functional methods do not allow even
approximate access to dynamical susceptibilities of correlated materials.
The static susceptibilities are essential to ensure the stability 
of the obtained solutions.

In this paper we study the dynamical susceptibilities of the
two-orbital Hubbard model on a bipartite lattice at half filling. In
particular, we focus on the mechanism of transition between the EC and
spin-state order (SSO) phases. The studied phase transitions involve
both continuous and discrete symmetry breaking and multi-atomic unit
cells. Besides understanding the physics of the model and assessing
the performance of the method, this work is the next step towards
similar investigations within the LDA+DMFT framework for real
materials.

\section{Computational Method}
The studied model Hamiltonian reads
\begin{equation}
\label{eq:model}
\begin{split}
  H = & \sum_{<ij>,\sigma} (t_a a_{i\sigma}^{\dagger} a_{j\sigma} + t_b b_{i\sigma}^{\dagger} b_{j\sigma}) + H.c. \\
      & +\frac{\Delta}{2}\sum_{i,\sigma}(n^a_{i\sigma}-n^b_{i\sigma}) \\
      & + U \sum_{i,\alpha}n^\alpha_{i\uparrow}n^\alpha_{i\downarrow}+
        \sum_{i,\sigma\sigma'}(U'-J\delta_{\sigma\sigma'}) n^a_{i\sigma}n^b_{i\sigma'},
\end{split}
\end{equation}
where $a^{\dag}_{i\sigma}$ and $b^{\dag}_{i\sigma}$ are the fermionic
creation operators for electrons in the respective orbitals $a$ and
$b$, with spin $\sigma$, at site $i$ of a square lattice. The first term
describes nearest neighbor hopping. The remaining terms, containing
the particle number operators $n^{c}_{i,\sigma} \equiv 
c^{\dag}_{i\sigma}c_{i\sigma}$, correspond to the crystal-field
$\Delta$, the Hubbard interaction $U$, and Hund's exchange $J$ 
in the Ising approximation.
The values $U=4$, $J=1$, and $U'=U-2J$ are fixed throughout this
study. The remaining parameters $t_{a}$, $t_{b}$, $\Delta$ as well as
the temperature $T$ are varied. All calculations reported here are
performed for the filling of two electrons per atom.


We follow the standard DMFT procedure, in which the lattice model is
mapped onto an auxiliary Anderson impurity model
(AIM)~\cite{Georges1992,Jarrell1992}. The AIM is solved numerically,
using the ALPS implementation~\cite{Bauer2011, Shinaoka2016,
  Gaenko2017} of the matrix version of the strong-coupling
continuous-time quantum Monte-Carlo (CT-QMC)
algorithm~\cite{Werner2006a}.

The model hosts several competing phases, which can be distinguished
by the mean values of operators
\begin{equation}
\label{eq:operators}
\begin{split}
\phi_i^\gamma  & =   R_i^\gamma+iI_i^\gamma=
  \sum_{\alpha\beta} \sigma^\gamma_{\alpha\beta}
  a_{i \alpha}^{\dagger}
    b_{i\beta}^{\phantom\dagger}\\
  O_{i} &= \sum_{\sigma} (n^{a}_{i\sigma} - n^{b}_{i\sigma}) \\
  S_{i}^z&=\sum_{c=a,b}(n^{c}_{i\uparrow} - n^{c}_{i\downarrow}).
  \end{split}
  \end{equation}
Here $\phi_i^\gamma$, with the Hermitean and anti-Hermitean parts $R_i^\gamma$ and $iI_i^\gamma$,
creates an $S=1$ exciton on site $i$. The $\sigma^{\gamma}$ ($\gamma=x,y,z$)
are Pauli matrices, which represent the spin polarization of the
exciton. With the density-density form of the interaction, which
mimics an easy-axis single-ion anisotropy,
$\langle\phi_i^z\rangle =0$ applies throughout the studied parameter
range~\cite{Kunes2015}. The $O_i$ and $S_i^z$ represent the local
orbital polarization and the $z$-component of the spin moment,
respectively.




The susceptibilities $\chi^{X}(\bk,\omega)$ are
obtained by analytic continuation~\cite{Gubernatis1991, geffroy2019}
of their Matsubara representations
\begin{equation}
\label{eq:susc}
\chi^{X}(\bk,i\nu_n)=
\!
\!
\int_0^{\beta}\!\!\!\!\!\mathrm{d}\tau 
e^{i\nu_n\tau} 
\expval{
  X_{-\bk}(\tau)X_{\bk}(0)}
-|\expval{X_\bk}|^2\\
\end{equation}
where the Fourier transform is defined as $X_{\bk}=\frac{1}{\sqrt{N}}\sum_\bR
e^{-i\bk\cdot\bR}X_\bR$.
The observables $X$ of interest are represented by the operators listed in
(\ref{eq:operators}).
\begin{figure}
    \includegraphics[height=0.7\columnwidth]{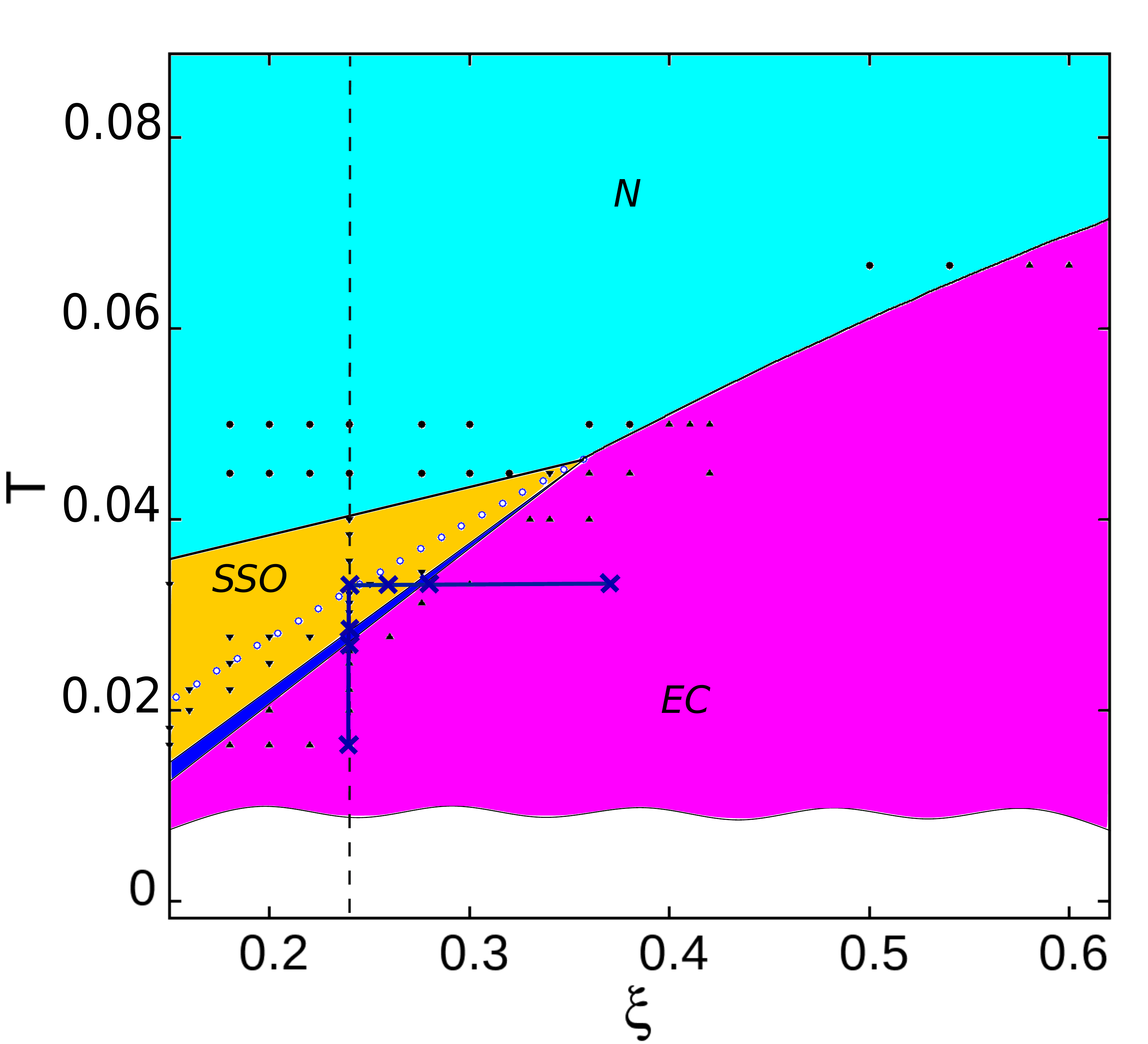}
    \includegraphics[height=0.7\columnwidth]{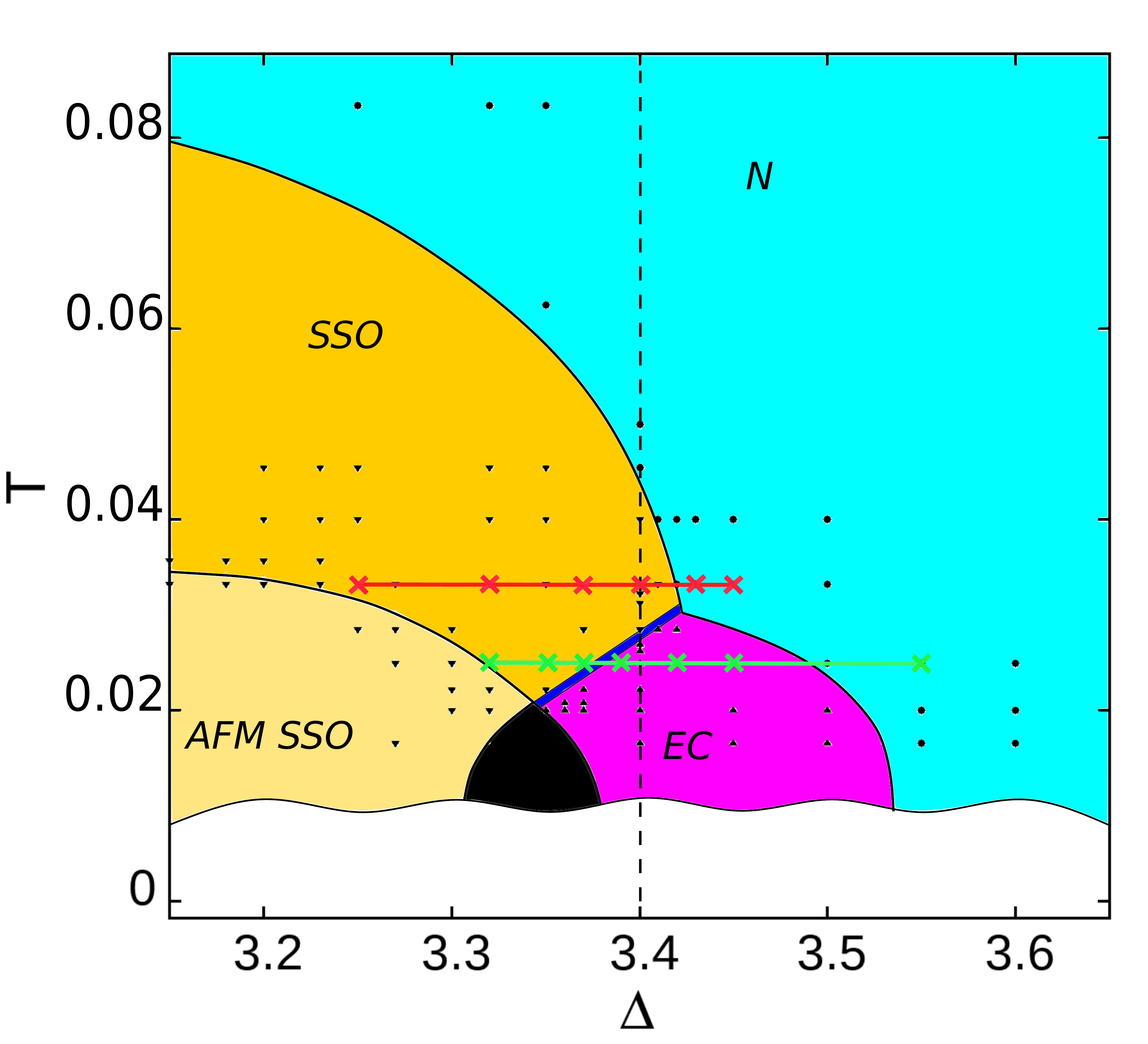}
        \caption{\label{fig:diag}
    Cuts through the phase diagram of the studied model along the
    $\xi$--$T$ ($\Delta=3.4$, top) and $\Delta$--$T$ ($\xi=0.24$,
    bottom) planes. Black symbols mark the parameters for which a
    calculation was performed. The dashed vertical line is common to
    both panels. The narrow blue wedge separating the EC and SSO
    phases represents the SS phase. The red, blue and violet cuts with
    crosses mark the points for which the susceptibilities are
    analyzed in Figs.~\ref{fig:susc1}, \ref{fig:susc2} and
    \ref{fig:susc3}, respectively. Open blue circles in the left panel
    were taken from Ref.~\onlinecite{Kunes2014a} and indicate
    instability of the normal (N) phase towards EC. The black wedge
    separating the AFM SSO and EC phases in the right panel indicate a
    putative coexistence regime accompanying a first order
    transition. (Actual calculations investigating this transition
    were not performed).}
\end{figure}

We start with the 1P propagators at 300 Matsubara frequencies to
obtain the bare susceptibilities (both local and lattice bubble
terms), which are then transformed into the Legendre polynomial
representation~\cite{Boehnke2011}.
The 2P correlation function is sampled using the CT-QMC algorithm. The
local 2P-irreducible vertex $\Gamma$ is obtained by
inverting the impurity Bethe-Salpeter
equation (BSE)~\cite{Georges1996, Kunes2011, vanLoon2015,Krien2017}.
Using this vertex to solve the lattice BSE, we obtain the lattice
correlation functions. This procedure is 
performed independently for each bosonic Matsubara frequency. We have
found that using 10 bosonic frequencies allows for a stable and good
quality analytic continuation. We use between 22 (for the zeroth bosonic frequency)
and 30 Legendre coefficients (for the ninth bosonic frequency). A
sizable reduction of the computational and storage cost can be
achieved with the procedure of Ref.~\onlinecite{SpM, shinaoka2020}.
 
The susceptibility $\chi^{X}(\bk,i\nu_n)$ is
a diagonal element of the particle-hole susceptibility matrix
$\boldsymbol{\chi}(\bk,i\nu_n)$ obtained by summation 
of the lattice correlation function over the Legendre
coefficients.
The matrix $\boldsymbol{\chi}(\bk,i\nu_n)$ is indexed by pairs of flavors
(spin/orbital/site) inside the unit cell, while the
inter-cell structure is diagonalized by going to
the reciprocal space. With four flavors per site, $\boldsymbol{\chi}(\bk,i\nu_n)$
has  dimension $4^2$ for 1-atom cell. In phases with
2-atom cells  $\boldsymbol{\chi}(\bk,i\nu_n)$ has the dimension
$(2\times 4)^2$. However, thanks to the locality of the 2P-irreducible
vertices, the BSE can be written in a closed form for elements of the
type $\chi_{ii,jj}$, where $i$, $j$ are the site indices. Therefore
the diagonal elements in (\ref{eq:susc}) can be obtained by working
with matrices of the flavor dimension  $2\times 4^2$, i.e., linear in
the number of sites per the unit cell.

To ensure comparability of $\chi^{X}(\bk,i\nu_n)$ in different
phases (various unit cells) we present all susceptibilities
(\ref{eq:susc}) in the large Brillouin zone of the 1-atom unit cell.
In the phases with 1-atom unit cell the susceptibility is diagonal
in $\bk$. In phases with $\sqrt{2}\times\sqrt{2}$ 2-atom unit cells
there are no-zero off-diagonal elements connecting $\bk$ and
$\bk+(\pi,\pi)$. The transformation from the 2-atom unit cell, in
which the BSE inversion is performed, is given by
\begin{equation}
\begin{split}
    \chi(\bk)&=\tilde{\chi}(\bk^\prime)_{11,11}+\tilde{\chi}(\bk^\prime)_{22,22}\\
           &+\exp(ik_y)  \tilde{\chi}(\bk^\prime)_{11,22}+
           \exp(-ik_y) \tilde{\chi}(\bk^\prime)_{22,11},
\end{split}
\end{equation}
where $\tilde{\chi}$ and
$\bk^\prime\equiv(k^\prime_x,k^\prime_y)=(k_y-k_x,k_y+k_x)$ are
related to the 2-atom unit cell. The subscripts of $\tilde{\chi}$
refer to the two sites in the 2-atom unit cell (The orbital and spin
indices are not shown for sake of simplicity).







\section{Results and Discussion}
\subsection{Phase diagram and order parameters}
In Fig.~\ref{fig:diag} we show the phase diagram of the model 
in the $\xi$--$T$ plane of band asymmetry parameter $\xi=\frac{2t_at_b}{t_a^2+t_b^2}$
and temperature $T$ at fixed crystal field $\Delta$, and in the 
$\Delta$--$T$ plane at fixed $\xi$.
The phase boundaries are obtained by combination of the calculated order parameters and diverging susceptibilities.
The phase diagram in Fig.~\ref{fig:diag}a generalizes that
of Ref.~\onlinecite{Kunes2014a} to the ordered phases.
The phase diagram in Fig.~\ref{fig:diag}b should be compared to the
phase diagrams of related strong-coupling models in
Refs.~\onlinecite{Kunes2015,Tatsuno2016}. Unlike previous
studies~\cite{Kunes2014a,Hoshino2016} where the
instabilities of the normal phase were investigated, here we perform
linear response calculations also in the thermodynamically stable ordered phases. Four distinct ordered phases are identified.
\begin{figure}[b]
    \includegraphics[width=0.498\columnwidth]{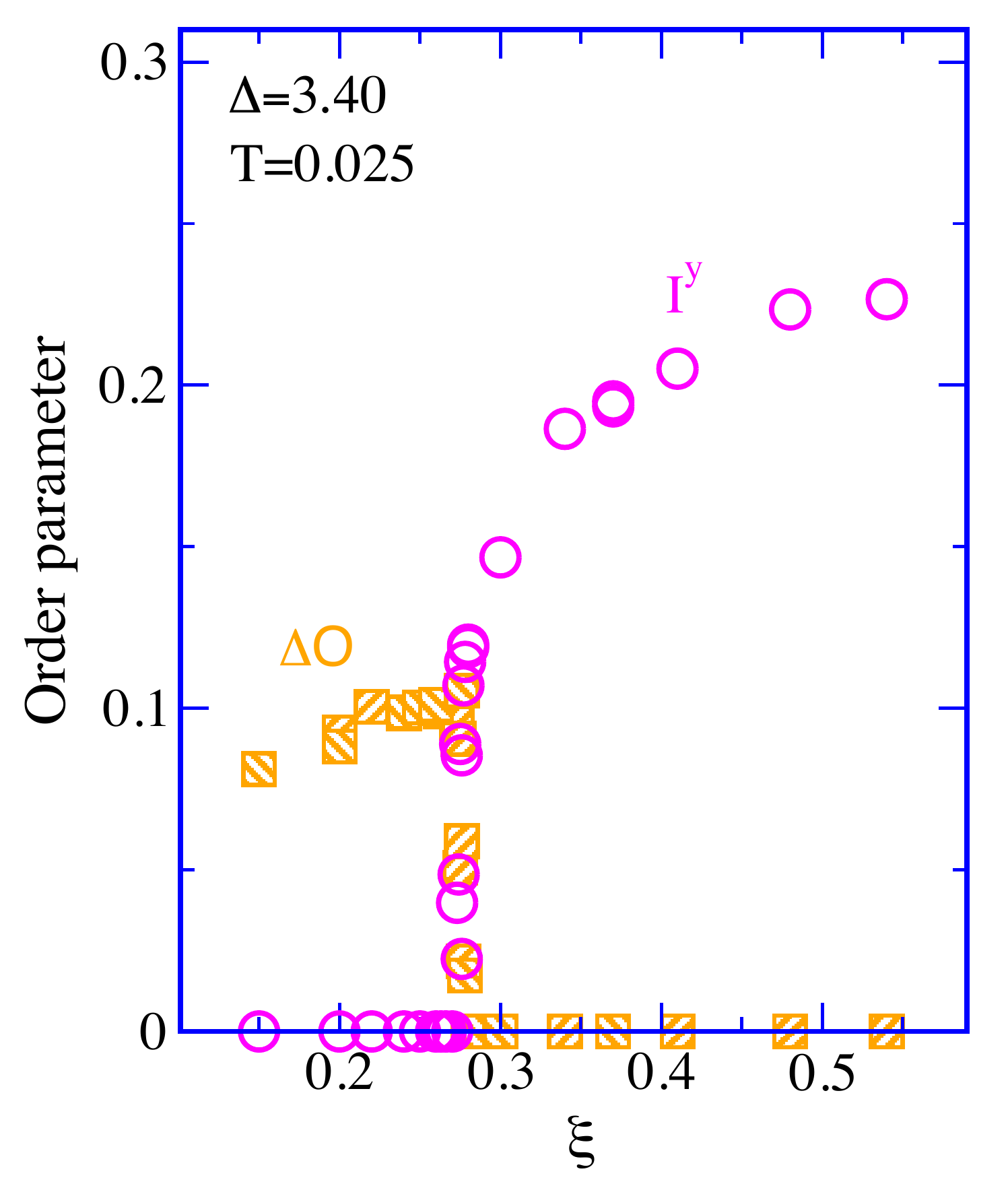}
     \includegraphics[width=0.47\columnwidth]{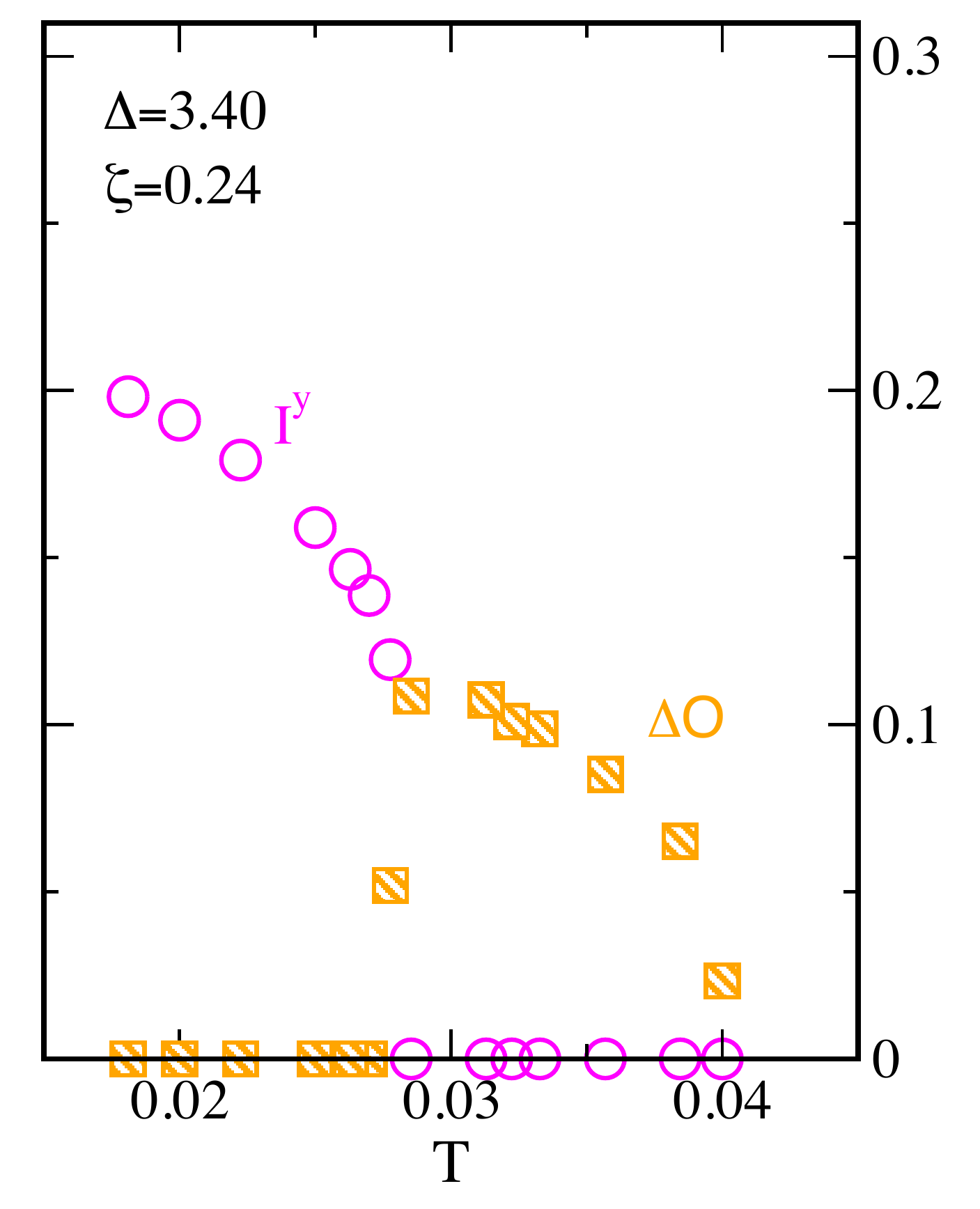}
    \caption{The order parameters: staggered orbital polarization
      $\Delta O$ and the uniform excitonic-condensate amplitude $\langle I^y\rangle$
      along the cuts in the phase diagram Fig.~\ref{fig:diag} marked
      by the frame colors.}
    \label{fig:order}
\end{figure}
\begin{figure}[b]
    \includegraphics[width=0.98\columnwidth]{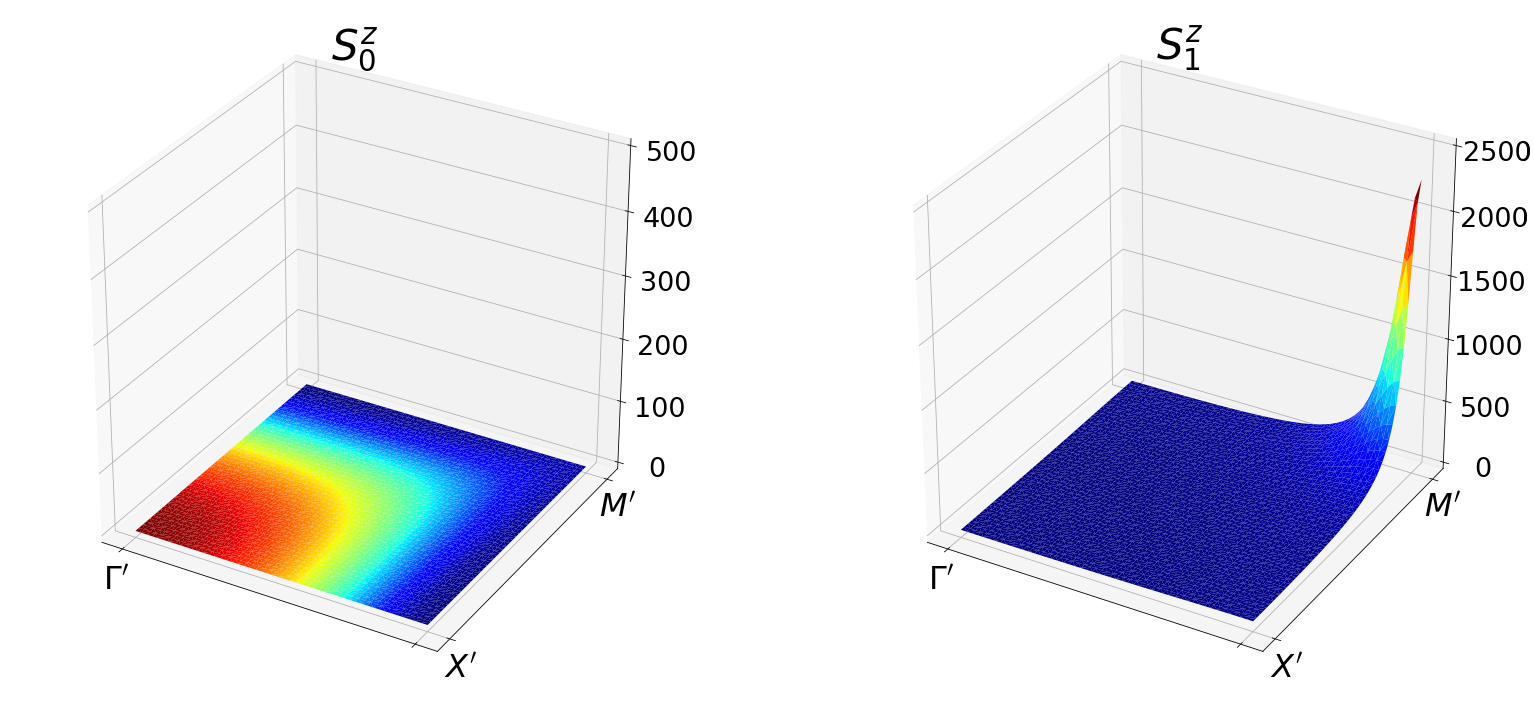}
    \caption{Site resolved static susceptibility
      $\chi^{S^z}(\bk^\prime)$ for $T=0.0333$, $\xi$=0.24, $\Delta$
      =3.23 in the SSO phase. It describes the response to a field
      applied only on the LS-like sublattice (left) and on the HS-like
      sublattice (right). The reciprocal vector $\bk^\prime$ is
      expressed with respect to the 2-atom unit cell.}
    \label{fig:susc_stat}
\end{figure}

{\it Polar excitonic condensate (EC).} This phase was analyzed in
detail in a number of previous
studies~\cite{Kunes2014c,Kunes2016,Geffroy2018,Nasu2016,Kaneko2014}. It is
characterized by a finite expectation value of $\langle \bphi_i
\rangle=\bphi$, which fulfills the
condition  ${ \bphi^* \times \bphi  =
  0}$~\cite{Balents2000b,Geffroy2018}. The EC phase preserves the
translation symmetry, but breaks two continuous $U(1)$ symmetries
associated with the global conservation of $\sum_iS_i^z$ and $\sum_iO_i$.
The EC order parameter 'lives' on a $T_2$ torus - it can
pick an arbitrary orientation in the spin $xy$-plane and an
arbitrary complex phase. Throughout the present study we fix its
orientation to $\langle I^y\rangle\neq 0$, while the other components
are zero.

{\it Spin state order (SS0).} The SSO phase in the two-band Hubbard
model was reported in Ref.~\onlinecite{Kunes2011} and in multi-orbital
material specific DMFT studies~\cite{Karolak15,Afonso2019}. It was
proposed as an explanation of high field experiments on
LaCoO$_3$~\cite{Altarawneh2012,Ikeda2016}. It is characterized by
staggered orbital polarization $\Delta O=(-1)^i\langle
O_i-\bar{O}\rangle$, where $(-1)^i$ describes the
$\sqrt{2}\times\sqrt{2}$ order and $\bar{O}$ denotes an average over
all lattice sites. The SSO is a strong-coupling effect that, unlike
the EC phase, does not have a weak-coupling
analog~\cite{Kunes2014a}. At $T=0$ the phase is a checkerboard
arrangement of LS and HS sites. In the studied parameter range
the LS-like sites are dominated by the LS state with a negligible HS
contribution. The population of the HS state on HS-like sites is only
up to 60\%, with the remainder being predominantly LS
states~\footnote{By occupation we mean the diagonal elements of the
  site-reduced density matrix operator.}.
The SSO phase breaks the translation symmetry, but
the continuous $U(1)$ symmetries associated with $S^z$ and $O$
conservation are preserved.

{\it Supersolid (SS).} The SS phase is characterized by the
simultaneous appearance of the EC and SSO orders~\cite{Boninsegni2012,Scalettar1995}. The SS phase breaks
all the symmetries broken by EC and SSO phases. We consistently find
very narrow strip of the SS phase at the boundary between the EC
and SSO phases, see Fig.~\ref{fig:order}.

{\it Antiferromagnetic spin state order (AFM-SSO)}. The SSO phase has
a large residual entropy associated with the spin disorder on the HS
sites. The nearest neighbor AFM exchange interaction
on the HS sublattice (3rd neighbor interaction on the original lattice) leads to a $2\times2$ order consisting in
checkerboard spin order on the HS sublattice. We did not actually
perform calculations in the AFM SSO phase, but determined the
SSO/AFM-SSO phase boundary as the divergence of
$\tilde{\chi}^{S^z}(M',0)$, see Fig.~\ref{fig:susc_stat}.
\begin{figure*}
\begin{minipage}{0.7\linewidth}
   \fcolorbox{green}{white}{\includegraphics[width=0.98\linewidth]{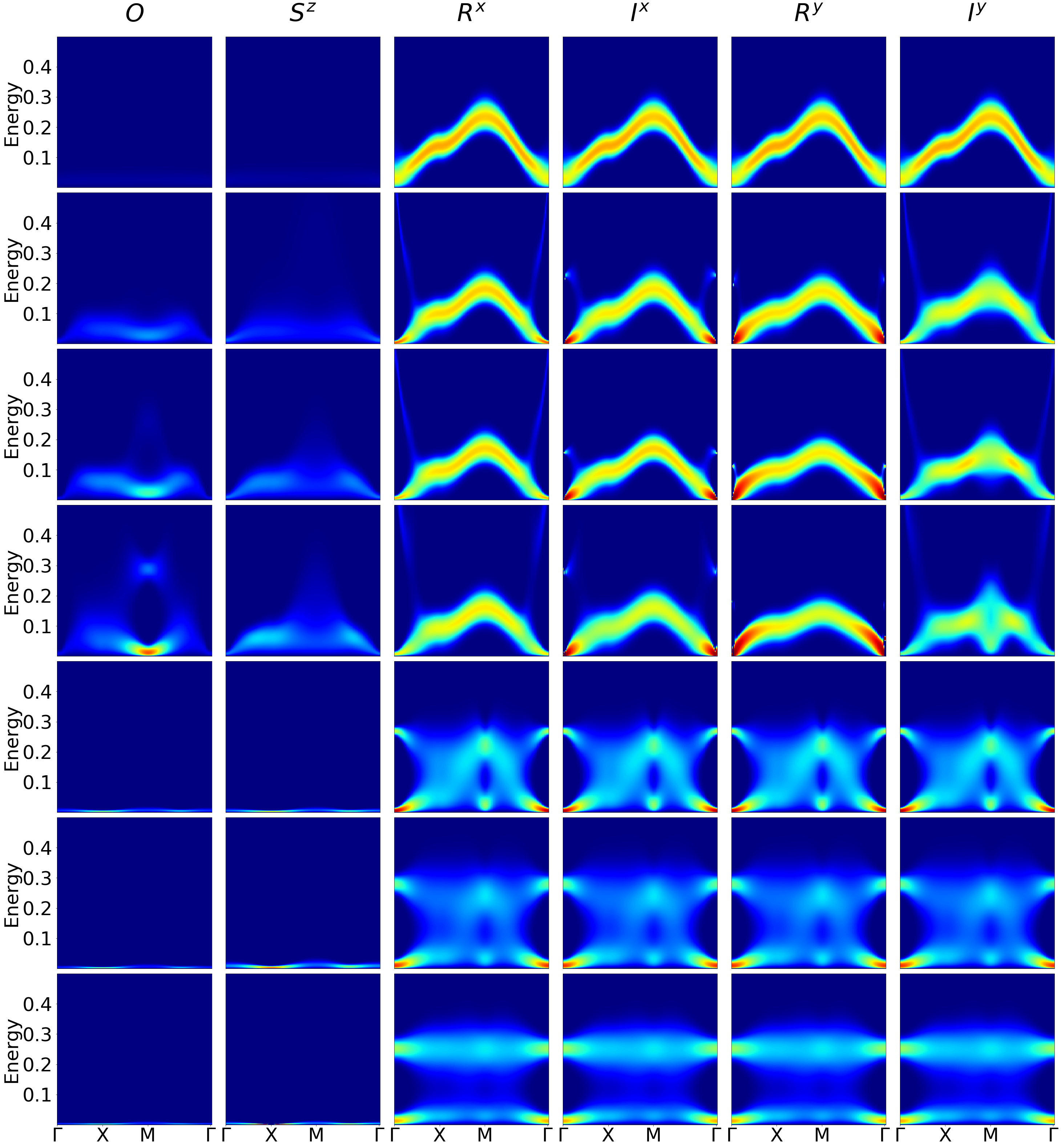}}
 \end{minipage}\hfill
 \begin{minipage}{0.28\linewidth}
    \caption{Evolution of the dynamical susceptibility
      $\chi^X(\bk,\omega)$ with the crystal field $\Delta$ along the
      green line ($\xi = 0.24$, $T=0.025$) in Fig.~\ref{fig:diag}b. The
      columns correspond to different Hermitean operators $X$, the
      rows correspond to different $\Delta$: 1) $\Delta=3.55$ normal
      phase, 2--4) $\Delta=3.45$, 3.42, 3.39 EC phase, 5--7)
      $\Delta=3.37$, 3.35, 3.32 SSO phase. The color coding represents
      the spectral density
      $B(\bk,\omega)=-\frac{1}{\pi}\operatorname{Im}\chi^X(\bk,\omega)$. In
      order to capture the entire dispersion in the presence of
      divergent density of the Goldstone modes we introduce a cut-off
      and plot $\frac{B}{B+\text{const}}$ with const=5.5. Note that in
      the SSO phase there is a large intensity of $\chi^{S^z}$ at
      $\omega\approx 0$ (difficult to see in the present figures)
      corresponding to large static response of local moments on HS
      sites.}
          \label{fig:susc1}
\end{minipage}
\end{figure*}

\subsection{Dynamical susceptibility}

\begin{figure*}
\begin{minipage}{0.7\linewidth}
    \fcolorbox{blue}{white}{\includegraphics[width=0.98\linewidth]{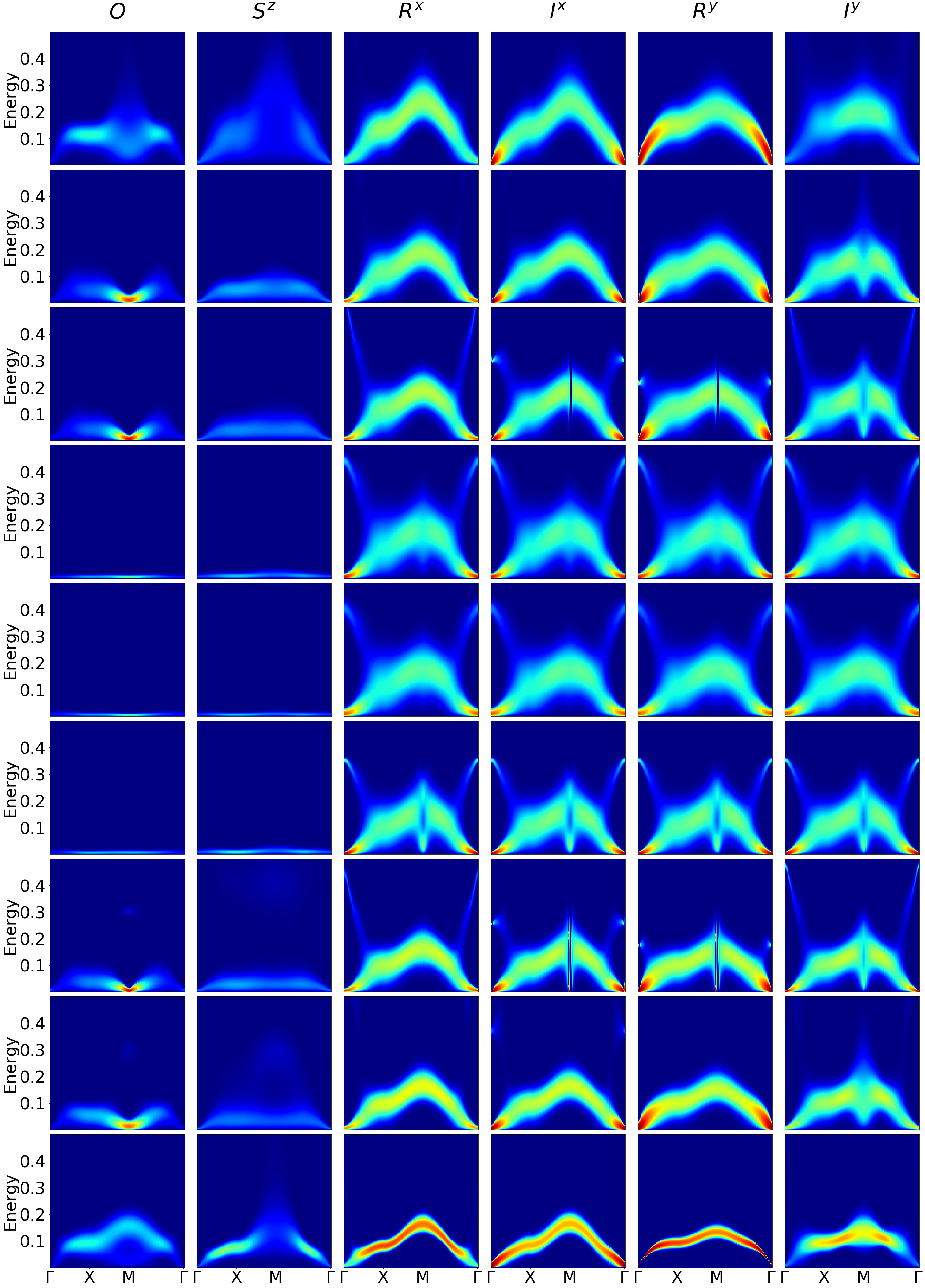}}
 \end{minipage}\hfill
 \begin{minipage}{0.28\linewidth}
    \caption{As in Fig.~\ref{fig:susc1}, evolution of the dynamical
      susceptibility $\chi^X(\bk,\omega)$ along the inverted L-shaped
      blue line in Fig.~\ref{fig:diag}a for $\Delta$ =
      3.4. The horizontal line corresponds to fixed temperature $T =
      0.033$, while the vertical line corresponds to fixed band
      asymmetry $\xi=0.24$. The rows correspond to different parameters:
      rows 1--2) $\xi=0.37$, 0.30; EC phase, 3) $\xi= 0.278$; SS
      phase, 4--5) $\xi= 0.26$, 0.24; SSO phase, 6) $T=0.0286$; SSO
      phase, 7) $T= 0.0278$; SS phase, 8--9) $T= 0.0370$, 0.0167; EC
      phase.}
     \label{fig:susc2}
\end{minipage}
\end{figure*}

\begin{figure}
   \fcolorbox{red}{white}{\includegraphics[width=0.98\columnwidth]{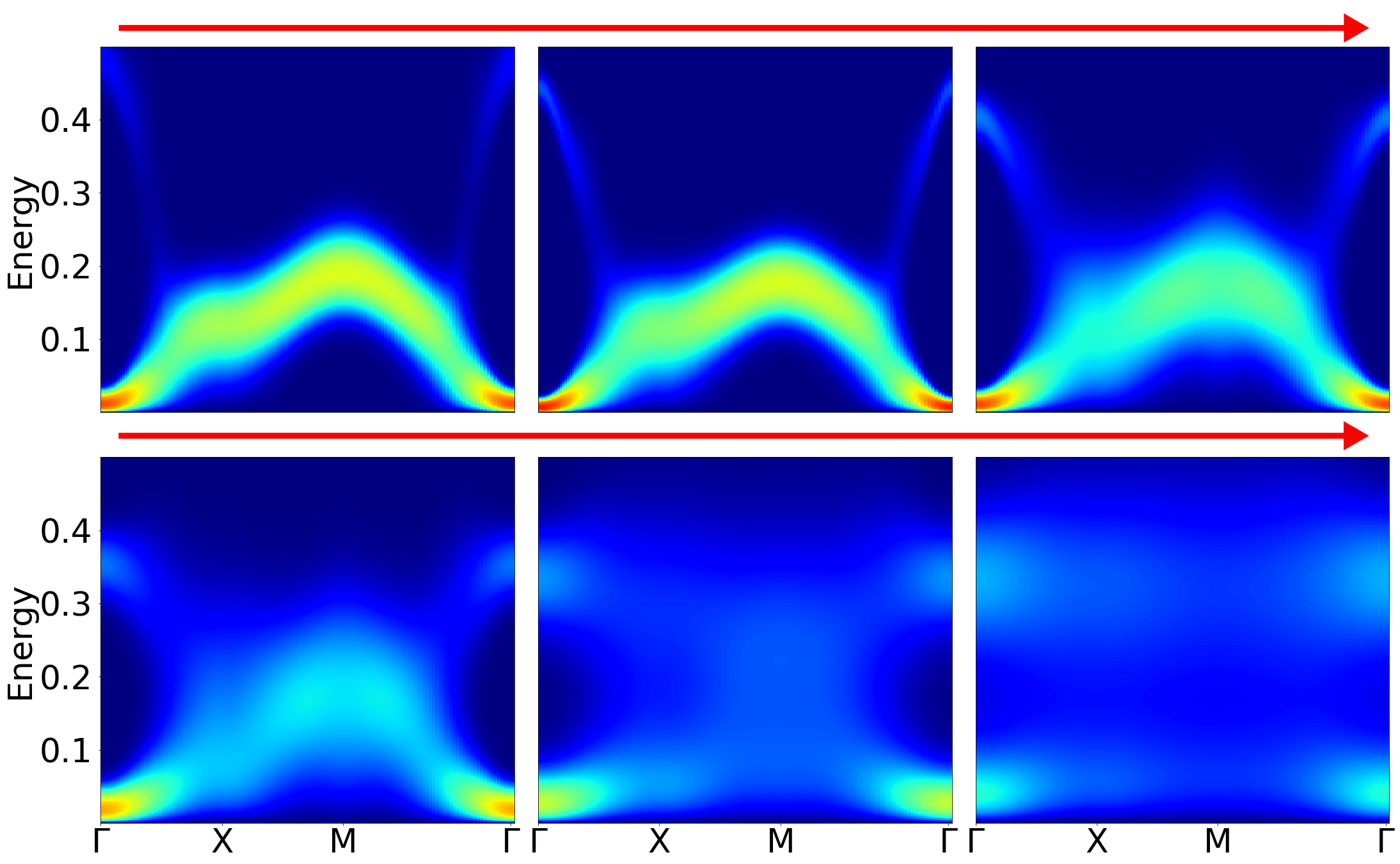}}
    \caption{Evolution of the dynamical excitonic susceptibility
      $\chi^X(\bk,\omega)$ with crystal field $\Delta$ along the red
      line($\xi$=0.24, $T=0.0333$) in  Fig.~\ref{fig:diag}b.1--2)
      $\Delta=3.45$, 3.42; normal phase, 3--6) $\Delta$ = 3.40, 3.37,
      3.32, 3.27; SSO phase.}
    \label{fig:susc3}
  \end{figure}
  \begin{figure}[b]
    \fcolorbox{blue}{white}{\includegraphics[width=0.98\columnwidth]{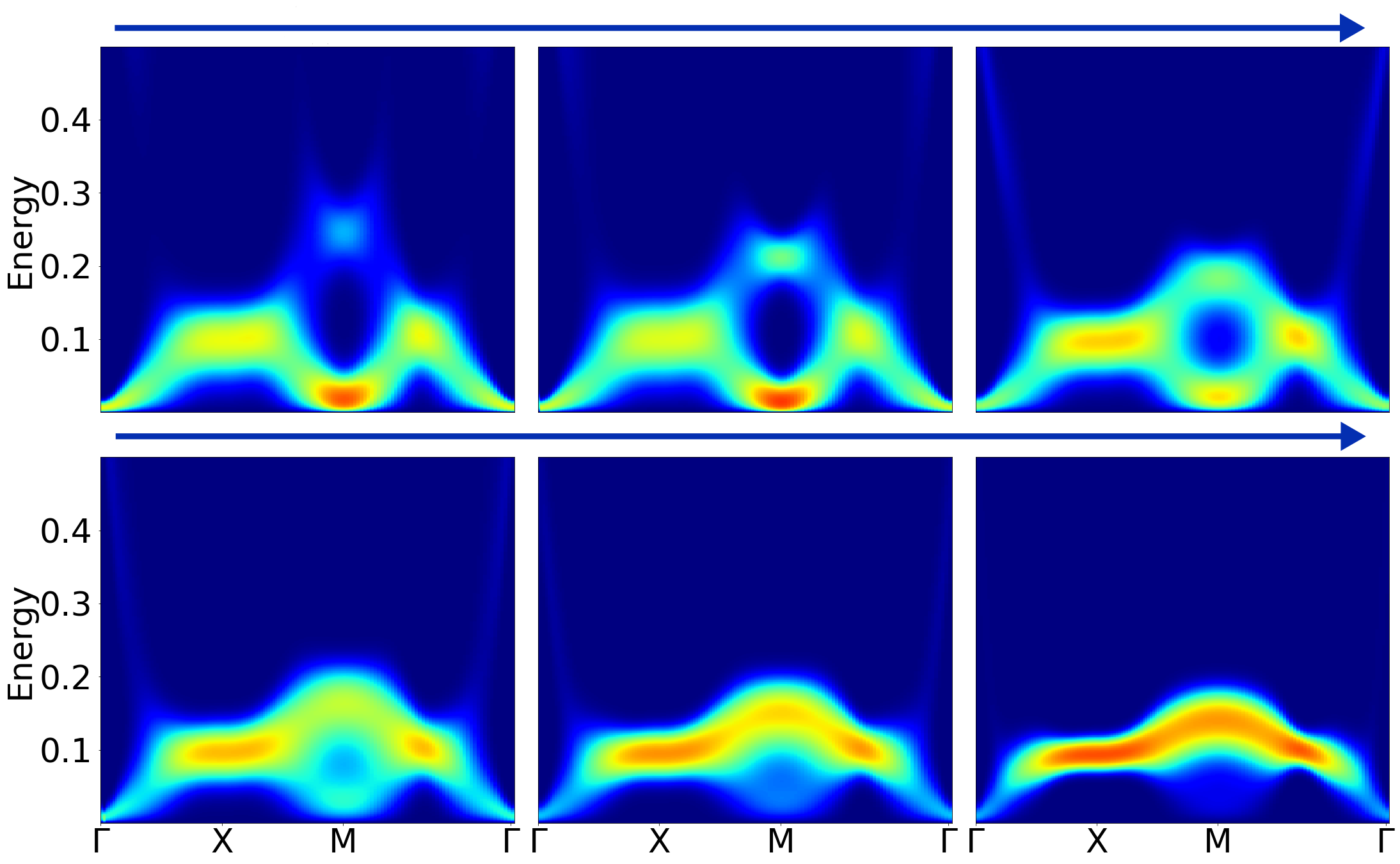}}
    \caption{Dynamical susceptibility $\chi^{\tilde{I}^y}(\bk,\omega)$
      in the EC phase of Fig.~\ref{fig:susc2} ($\Delta=3.40$;
      $\xi$=0.24; $T=0.0270$, 0.02632, 0.025, 0.0222, 0.02, 0.0167 in
      the order marked by the arrow) in the basis of the eigenmodes of
      the static susceptibility.
     \label{fig:ECeig}}
\end{figure}
\begin{figure}
   \fcolorbox{green}{white}{\includegraphics[width=0.8\columnwidth]{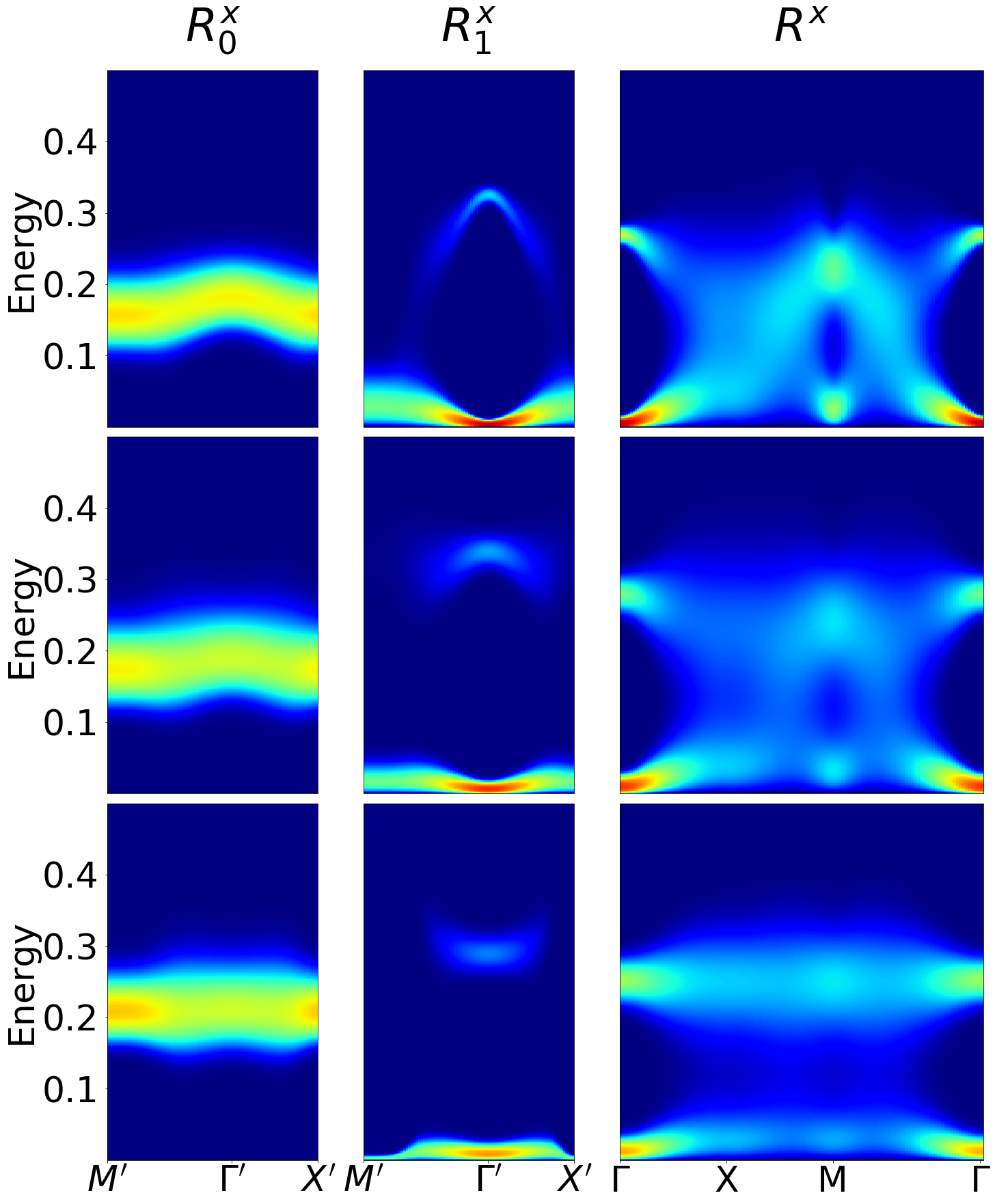}}
    \caption{Dynamical excitonic susceptibilities $\chi^X(\bk,\omega)$
      in the SSO phase of Fig.~\ref{fig:susc1} ($T=0.025$;
      $\Delta=$3.37, 3.35, 3.32 from top to bottom) in the basis of
      the eigenmodes of the static susceptibility. The $R^x_0$ and
      $R^x_1$ columns refer to the two eigenmodes in the 2-atom unit cell
      of the SSO. The $k$-path refers to the Brillouin zone for the
      2-atom unit cell. The $R^x$ column reproduces the data from
      Fig.~\ref{fig:susc1} for comparison ($k$-path in the Brillouin
      zone of 1-atom unit cell).}
    \label{fig:SSO40eig}
\end{figure}
\begin{figure}
   \fcolorbox{red}{white}{\includegraphics[width=0.8\columnwidth]{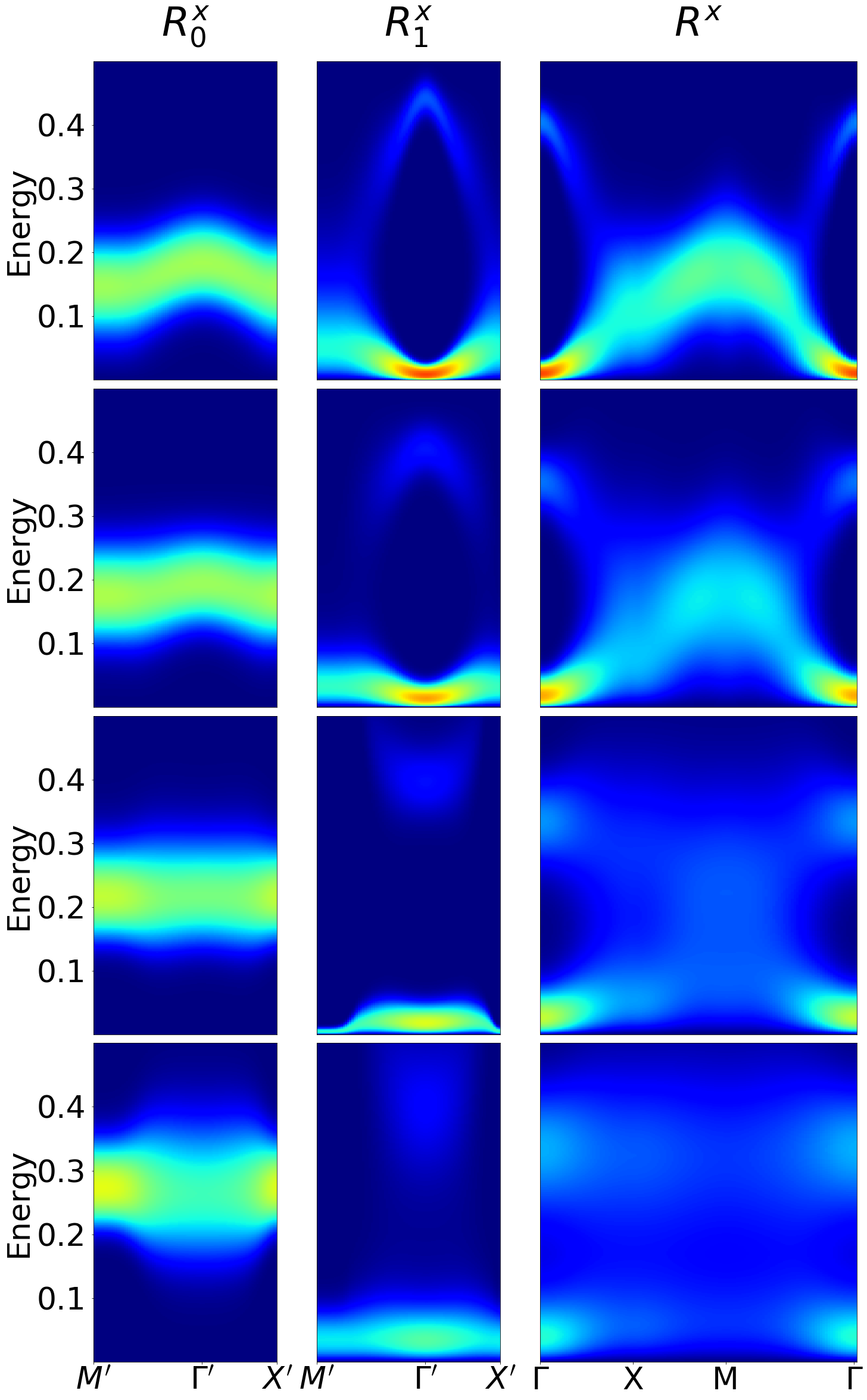}}
    \caption{Dynamical excitonic susceptibilities $\chi^T(\bk,\omega)$
      in the SSO phase of Fig.~\ref{fig:susc3} ($T=0.0333$;
      $\Delta=3.4$, 3.37, 3.32, 3.25 from top to bottom) in the basis
      of the eigenmodes of the static susceptibility. The notation is
      the same as in Fig.~\ref{fig:SSO40eig}.}
    \label{fig:SSO30eig}
\end{figure}

The main focus of this work is the behavior of the dynamical
susceptibility across the transition between the EC and SSO
states.
In Fig.~\ref{fig:susc1} we show the evolution of $\chi^X(\bk,\omega)$
($X$=$O$, $S^z$, $R^x$, $I^x$, $R^y$, $I^y$) along the $\Gamma$-X-M-$\Gamma$ 
path in the 2D Brillouin zone with increasing crystal fields $\Delta$.

First, we review the discussion of the N--EC transition from
Ref.~\onlinecite{geffroy2019}. The normal (N) phase is characterized by gapped
excitonic dispersion, reflected in all excitonic susceptibilities. The
equivalence of $x$ and $y$ elements originates from the $S^z$-conservation,
while the equivalence  of $R$ and $I$ elements originates from
the $O$-conservation. The $O$- and $S^z$-susceptibilities exhibit no
dynamics (non-zero only for $\nu_n=0$) and vanish at low
temperature. Reducing $\Delta$ results in closing of the excitonic
gap and eventually transition to the EC phase, where the equivalence
of excitonic susceptibilities is broken.
Deep in the EC phase we can distinguish $x$ and
$y$ excitonic modes with distinct dispersion. The corresponding
susceptibilities $R_x$, $I_x$ and $R_y$, $I_y$ follow these
dispersions, but have vastly different amplitudes at low energies.
The $I^x$ and $R^y$ exhibit linear dispersion and diverging amplitudes
at $\Gamma$, reflecting the spin-rotation and phase-rotation Goldstone
modes~\cite{geffroy2019}.

The $S^z$- and $O$-susceptibilities acquire non-zero dynamics due to
the $S^z$--$R^x$ and $O$--$I^y$ coupling in the EC
phase. The induced dynamics of $S^z$ was explained in terms of the
strong-coupling model in Ref.~\onlinecite{geffroy2019}, see also SM~\cite{SM}.
The dynamical response of $O$ can be understood along similar lines.
In the strong coupling limit $O_i$ maps onto the number operator of excitons
${O_i=d_{ix}^{\dagger}d_{ix}^{\phantom\dagger}+d_{iy}^{\dagger}d_{iy}^{\phantom\dagger}}$.
Replacing $d_{iy}^{\dagger}$ with $i\tfrac{\phi}{2}+d_{iy}^{\dagger}$
in the EC phase we find ${O_i\sim -\tfrac{\phi}{2}(d_{iy}^{\dagger}-d_{iy}^{\phantom\dagger})}$,
and thus the correlation function of $O$ follows that of $I^y$. 
For a more rigorous derivation see SM~\cite{SM}.
We point out that all the above identifications are understood relative to the
orientation of the EC order parameter: $\langle I^y\rangle\neq 0$.

As we near the SSO phase the behavior of the $O$, $S_z$ and $I_y$
susceptibilities changes qualitatively. The $O$ and $S_z$ dynamics
cease to be slave to the excitonic dynamics and their dispersions stop
to follow the excitonic ones. Similar behavior is observed as we
approach the phase boundary as a function of crystal field $\Delta$,
Fig.~\ref{fig:susc1}, band asymmetry $\xi$ or temperature $T$,
Fig.~\ref{fig:susc2}. The $O$ susceptibility develops a hot spot at
the $M$ point, a precursor of the SSO phase, which is accompanied by softening of
$\chi^I_y$ at $M$. 
Similar behavior at the $M$ point was previously observed at zero temperature for spinless
hard-core bosons on square lattice and interpreted as roton
excitations known from superfluid helium~\cite{Scalettar1995}. We
provide the strong-coupling mean-field analysis of the softening and
EC--SSO transition in the Supplemental Material~\cite{SM}.

The demise of the EC phase due to the softening of the excitonic mode
accompanied by the divergence of ${\chi^O(\bk=M,0)}$ opens the
possibility for a continuous transition between the EC and
SSO phases via an intermediate SS
phase. Indeed, we find several solutions with both EC and SSO order,
Fig.~\ref{fig:order}, which fall into an narrow strip of
parameters. We point out that a similar situation was found in
Ref.~\onlinecite{Scalettar1995}.

In the SSO phase, we observe the remains of broad excitonic
dispersion in the vicinity of the phase boundary. It is important to 
point out that
at the studied temperatures, the
LS sites 
host almost exclusively the LS state, but the HS sites host
still up to 75\% LS and only 25\% HS states~\cite{Kunes2011}.
Proceeding deeper into the SSO phase the excitonic dispersion gives way
to two almost flat bands. These can be associated with creation of an exciton
(LS to HS transition) on the LS site (upper band) and annihilation of an exciton
(HS to LS transition) on the HS site (lower band).

\subsection{Mode analysis}
The connection between dynamical susceptibilities in
Figs.~\ref{fig:susc1}, \ref{fig:susc2} and \ref{fig:susc3} on the one
hand, and bosonic dispersions obtained in the strong coupling 
model~\cite{SM,Scalettar1995,Nasu2016} on the other hand, is not
straightforward. In the strong-coupling limit and at $T=0$, the
susceptibilities follow the dispersions of the $d_x$ or $d_y$ bosons
with intensity depending on the specific correlation function. Our
model is not in the strong-coupling limit and partly falls into a high
temperature regime.
The bosonic modes, a 2P basis in which $\boldsymbol{\chi}(\bk,\omega)$
is diagonal~\footnote{Unless symmetry is enforced, the diagonal
  character can be  only approximate as it is not possible to
  diagonalize $\boldsymbol{\chi}(\bk,\omega)$ for all bosonic
  frequencies simultaneously},
are not immediately obvious.

We attempt to obtain approximate modes by diagonalizing the static
susceptibility $\boldsymbol{\chi}(\bk,\omega=0)$. This procedure is
trivial in the normal phase, because each of the four mutually equal
excitonic susceptibilities forms a diagonal block of
$\boldsymbol{\chi}(\bk,\omega)$. In the SSO phase the excitonic
susceptibilities do not mix with other elements of
$\boldsymbol{\chi}(\bk,\omega)$ or with each other, and the
diagonalization is reduced to $2\times2$ blocks spanned by the two
sites of the 2-atom unit cell.

The dominant effect of diagonalization in the EC phase is to combine
$I^y$ with $O$ into a single low-energy mode with large spectral weight
$\tilde{I}^y$. In Fig.~\ref{fig:ECeig}, we show the evolution of
$\chi^{\tilde{I}^y}(\bk,\omega)$ as we approach the EC/SS phase
boundary by varying the temperature along the cut analyzed in
Fig.~\ref{fig:susc2}. Deep in the EC phase, $\chi^{\tilde{I}^y}$ is
essentially identical to $\chi^{{I}^y}$. As we get closer to the phase
boundary, a mode softening at the $M$-point is clearly
observable. Interestingly, it does not proceed as a smooth deformation
of the dispersion curve captured by the strong-coupling
model~\cite{SM}, but rather through a spectral weight transfer between
the upper and low branch of the O-ring structure observed at
$T$=0.025. We attribute this behavior to the finite temperature.

In Figs.~\ref{fig:SSO40eig},~\ref{fig:SSO30eig}, we show the mode
decomposition in the SSO phase. For all parameters we find two bands
reminiscent of the strong-coupling
behavior~\cite{Scalettar1995,Nasu2016}. Deep in the SSO phase the
bands are flat. The lower one corresponds to eliminating a HS exciton
on the nominally HS sublattice. The upper band corresponds to creating
a HS exciton on the nominally LS sublattice. With increasing crystal
field $\Delta$, the character of the bands changes and they become
dispersive, while the gap between them shrinks. This behavior is
somewhat counter-intuitive, since the difference of HS and LS energies
in an isolated atom follows an opposite trend. The explanation lies
with the nearest-neighbor repulsion between HS excitons. Increasing $\Delta$ causes
a decrease in concentration of HS states on the nominally HS
sublattice, which reduces the HS-HS repulsion that has to be overcome
when creating a HS exciton on the nominally LS
sublattice. Simultaneously, a minute shift of the lower excitonic band
leads to a condensation as the SSO/SS boundary is approached.
At the SSO/N boundary, Fig.~\ref{fig:SSO30eig}, the temperature is too high
for the excitons to condense. We observe a complete closing of the gap between the two
bands, which become a back-folded image of the excitonic band from the
1-atom unit cell.
In addition to the two main bands, we observe a weak high-energy
feature around the $\Gamma'$-point, which does not have a
strong-coupling $T=0$ counterpart. This feature exhibits a rather
strong dispersion and it is most pronounced close to the boundary of
SSO with either the SS or normal phases,
Figs.~\ref{fig:SSO40eig},~\ref{fig:SSO30eig}~\footnote{Note that
  there is a minor mismatch between the energy of this feature
  obtained in the 2nd and 3rd column of these figures. We attribute
  this to analytic continuation procedure, which is performed for the
  different bases independently and may have difficulty with accurate
  positioning of small high-energy peak in the spectrum containing
  large low-energy peak.}. 

\section{Conclusions}
We have studied the dynamical susceptibility across several phase
transitions in the two-orbital Hubbard model using DMFT. We have
observed a narrow slip of supersolid phase separating the spin-state order from 
the excitonic condensate. Approaching the spin-state ordered phase from the
exciton condensate is heralded by the softening of a specific collective
mode at the $M$-point of the Brillouin zone, identified as the roton
instability in Ref.~\onlinecite{Scalettar1995}. At low temperatures
the spin-state ordered phase removes the spin degeneracy by developing
antiferromagnetic order with $2\times 2$ periodicity. 

The present calculations demonstrate the utility of linear response
DMFT formalism for understanding complicated phase diagrams and phase
transitions involving the breaking of both discrete and continuous
symmetries. While the DMFT susceptibilities in the studied parameter
range qualitatively agree with the strong-coupling generalized
spin-wave treatment~\cite{Scalettar1995,Nasu2016,SM}, they contain
features that are beyond this description. Last but not least, we have
shown that the symmetry breaking in the exciton condensate gives rise
to dynamical response in the spin- and orbital-density channels. These
may be studied by standard experimental probes such as inelastic x-ray
or neutron scattering, which do not couple directly to the spin-triplet
excitonic channel.

\begin{acknowledgements}
The authors thank A. Kauch and R. T. Scalettar for comments and
critical reading of the manuscript. This work was supported by the
European Research Council (ERC) under the European Union's Horizon
2020 research and innovation programme (grant agreement
No.~646807-EXMAG). The authors acknowledge support by the Czech
Ministry of Education, Youth and Sports from the Large Infrastructures
for Research, Experimental Development and Innovations project
„IT4Innovations National Supercomputing Center – LM2015070“.
\end{acknowledgements}

\bibliography{supersolid}
\end{document}